A color-corrected, high-contrast catadioptric relay for high-resolution biological photolithography

Timm Michel[1,2], Jürgen Behr[2], Hamed Sabzalipoor[2], Gisela Ibáñez-Redín[3], Jory Lietard[3], Thomas Schletterer[4], Max Funck[5], Mark M. Somoza[2,3,6] *

[1]Technical University of Munich, Germany; TUM School of Life Sciences.

[2]Leibniz Institute for Food Systems Biology at the Technical University of Munich, Lise-Meitner-Straße 30, 85354 Freising, Germany

[3]Institute of Inorganic Chemistry, University of Vienna, Josef-Holaubek-Platz 2, 1090 Vienna, Austria

[4]Ingenieurbüro Thomas Schletterer, Homberger Ring 24, 07646 Stadtroda, Germany

[5]Funck Optics, August-Bebel-Straße 26-53, 14482 Potsdam, Germany

[6]Chair of Food Chemistry and Molecular Sensory Science, Technical University of Munich, Lise-Meitner-Straße 34, 85354 Freising, Germany

* Corresponding author: m.somoza.leibniz-lsb@tum.de

**Abstract**

Large-scale synthesis of DNA and RNA is a crucial technology for modern biological research ranging from genomics to nucleic acid therapeutics and for technological research ranging from nanofabrication of materials to molecular-level writing of digital data. Maskless Array Synthesis (MAS) is a versatile and efficient approach for creating the required complex microarrays and libraries of DNA and other nucleic acids for these applications and, more generally, for the synthesis of sequence-defined engineered and biological oligomers. MAS uses digital photomasks displayed by a digital micromirror device (DMD) illuminated by an appropriate light source and imaged into a photochemical reaction chamber with an optical relay system. Previously, Offner relay systems were used for imaging, but modern DMD formats with more and smaller micromirrors favor a different solution. We present a desktop MAS optical system with the larger numerical aperture and larger field of view required by 1080p and other large-format DMDs. The resulting catadioptric relay is well suited to modern DMDs in this application, and is corrected for first order axial and lateral color, enabling the use of high-power LEDs as inexpensive and long-lasting light sources spanning the ultraviolet-to-violet to perform the required photochemistry. Additional characteristics of the system, including high contrast and low scatter, make it ideal for reducing the error rates in photochemical synthesis of biomolecules.

**Introduction**

DNA microarrays or DNA "chips" have been important tools in genomics research and medical diagnostics for the last three decades[1-8]. The term "chip" was applied to DNA microarrays following the development of a photolithographic synthesis technology analogous to that used for integrated circuits or "microchips"[9]. Affymetrix introduced the first DNA microarrays based on modified I-line manufacturing approaches used by the semiconductor industry[10-12]. Microarrays are regular arrays of short oligomers (usually 30 to 80 nucleotides) attached to a flat substrate. Typical applications of microarrays have been, for example, whole genome gene expression analysis and comparative genomic hybridization (CGH), and thus required one or more DNA probes for each of the tens of thousands of genes present in animals and plants[11], or genomic DNA elements[13]. This and other applications, such genome-wide single nucleotide polymorphism arrays[14-15], arrays to determine the binding specificities of transcription factors and other DNA-binding proteins[16-19], and spatial transcriptomics arrays[20-21] used, e.g., in developmental biology, in cancer research and in food systems biology, require hundreds of thousands to several million unique DNA probes on each array, and therefore chemical-synthetic approaches able to achieve features with a pitch of some tens of microns or less. This pitch range is well within the capabilities of photolithography, but unlike integrated circuit manufacture, applications in DNA synthesis cannot make use of wavelengths much below 365 nm in order to achieve higher spatial resolution, due to absorption and photodamage to the DNA itself [22]. Photolithographic maskless array synthesis of DNA makes use of a modified form of the highly efficient phosphoramidite chemistry commonly used for oligonucleotide synthesis[23-24]. The building blocks, or monomers, for this type of synthesis are called phosphoramidites and each successive coupling reaction with these molecules elongates the nascent DNA strand by one adenosine (dA), cytidine (dC), guanosine (dG), or thymidine (dT) nucleotide. Although DNA synthesis for genomics applications has been the primary focus in microarray synthesis, recent years have seen the introduction of new DNA-based technologies that also require large-scale microarray-based synthesis[25], as well as applications requiring the analogous synthesis of RNA[26-31], non-canonical natural nucleic acids[32-33], chemically modified and engineered nucleic acids[34-39], non-nucleosidic sequence modifiers[40-43], as well as oligopeptides[44-45]. The key modification for light-directed synthesis is the use of photolabile protecting groups on the hydroxyl group of the 5' or 3' carbons of the pentose sugar of the DNA phosphoramidites. These protecting group prevents the coupling reaction unless it has been previously removed by light exposure. This synthesis scheme is shown schematically in **Figure 1**. A planar substrate, typically glass, is functionalized[46-52] to provide reactive groups for an initial coupling of a DNA phosphoramidite with a photolabile group, dT in this case. Spatial control of polymerization is accomplished by selectively illuminating pixels on the surface with 365 nm light. This photodeprotection is followed by a coupling reaction with one of the four DNA monomers and many cycles of exposure and coupling reaction result in a complex array of DNA oligonucleotides.

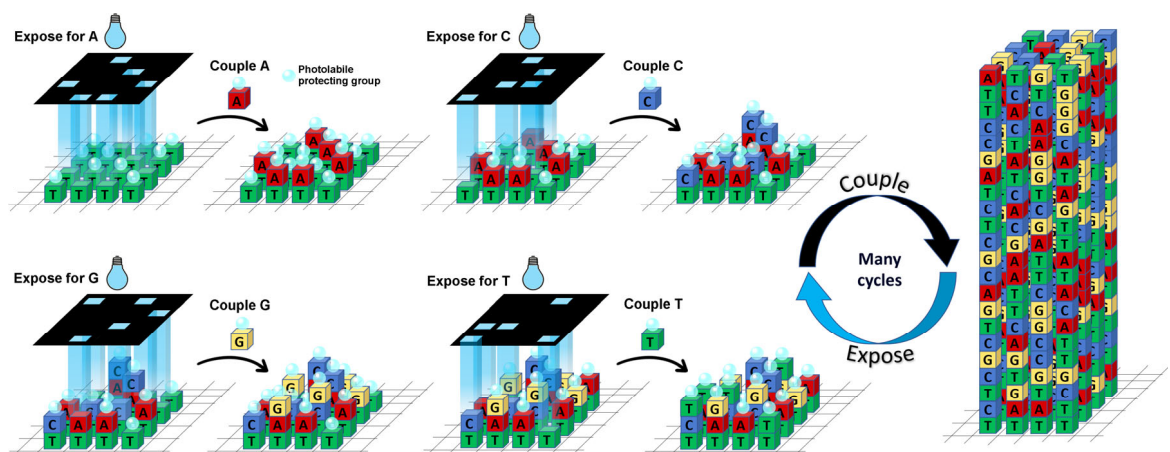

**Figure 1**. Schematic of the light directed synthesis. In the first reaction cycle, the dT monomer couples to reactive groups everywhere on the surface. Successive light exposures selectively remove photolabile groups present on the previously coupled DNA monomers. Many such cycles of coupling reactions followed by patterned exposure of the surface results in a complex array of DNA oligonucleotides.

A key innovation in the photolithographic synthesis of DNA microarrays was the insight that the then newly invented Digital Micromirror Devices by Texas Instruments[53] could be used to replace the chrome masks, and that the DMD, within an appropriate optical system, could lead to a simple and robust tabletop synthesis system[54]. This approach, termed maskless array synthesis or MAS, allows for fully digital control of the synthesis of complex DNA microarrays at both an industrial scale and in standard laboratory settings. The original maskless array synthesizers were designed around the Super Video Graphics Array (SVGA; 600 × 800; 10 × 14 mm) DMD with 480,000 16 µm pitch micromirrors. In this context, the optimal imaging system was found to be a 0.08 NA Offner relay yielding a spatial resolution of 2.7 µm using the 365 nm I-line of a mercury lamp[54]. The Offner relay[55-56] was a practical choice as it consists of only two spherical mirrors and provides sufficient open space for required DMD illumination from 24° of the normal. The relatively low resolution could be justified by size of the DMD micromirrors as well as the limited resolution (2 µm) of the best microarray scanners at the time[57]. In addition, the 0.08 NA reduces the relay cost and also reduces the amount of scattered light (~$NA^2$). Unlike semiconductor photolithography, where photoresist use mitigates most of the detrimental effects of stray light and reduced contrast on the patterning accuracy, the patterning in biological photolithography occurs by direct removal of photolabile protecting groups, which follows first order chemical kinetics. Therefore, stray light disproportionally increases the oligomer sequence error rate—specifically insertion errors—when photolabile protecting groups are removed on areas of the surface that are not directly exposed, enabling unintended reactions in the next coupling reaction.

Advances in DMD technology have resulted in larger arrays with smaller mirror size. These larger arrays enable extremely efficient parallel synthesis of nucleic acids, enabling large-scale bioanalytical assays relevant to recent advances in, e.g. RNA therapeutics and CRISPR-Cas gene editing, as well as new technologies such as digital information storage in DNA[58-64]. However, taking advantage of these new DMDs, such as the 0.95" 1080p array (1920 × 1080) with 2,073,600 10.8 µm pitch mirrors, or the 1.38" 4K array (4096 × 2160) with 8,847,350 mirrors with a 7.3 µm pitch, requires a new optical system with improved spatial resolution and a larger field of view. Here we describe an optical system custom designed for ultra-large-scale maskless synthesis of oligomers using modern DMDs. Optimized as a tabletop device for routine lab use, the new optical system has a numerical aperture of 0.12, corresponding to a spatial resolution of 1.5 µm at 365 nm, and has a field of view sufficient for diffraction-limited imaging of large-format DMDs up to 4K.

**Optical Design**

The primary design criteria for the new optical system for the relay were: (1) a numerical aperture of 0.12 to reach a spatial resolution of 1.5 µm at 365 nm, (2) a diffraction-limited field of view at least as large as the 1.38" 4K DMD, high optical contrast/low scatter to minimize insertion error rates, and sufficiently achromatic for illumination with both 365 nm and 405 nm LEDs, which have typical spectral widths of about 10 nm (FWHM) and relatively long red tails. In addition, the illumination system was specified to provide a 1080p DMD with an irradiance greater than 100 mW/cm$^2$ at 365 nm and to illuminate each mirror within 5 % of the mean irradiance.

**Optical relay**

The first generation of maskless array synthesizers (MAS) use a 0.08 NA (2.4 µm spatial resolution at 365 nm) Offner relays. Offner relays are simple, consisting of a primary concave spherical mirror with twice the radius of curvature of a smaller, convex, secondary mirror, which also defines the Fourier transform plane of the imaging optics. Further advantages of Offner relays is that they are achromatic and offer substantial clearance in front of the image and object plane, allowing the required frontal illumination of the DMD as well as space for the photochemical reaction cell and associated positioning mechanics.

The 0.08 NA Offner relay of the original MAS (**Figure 2a**) has a diffraction-limited field of view sufficient for the original SVGA DMD and the subsequent XGA DMD of the same dimensions, and marginally acceptable for the 1080p DMD (**Figure 2b**). Because of the advantages of Offner relays, we explored the possibility to using a 0.12 NA Offner relay (**Figure 2c**) to enable better resolution of the smaller mirrors of the 1080p and larger format DMDs. However, this NA increase requires a large increase in the size of the mirrors, as results in a diffraction-limited imaging only within a thin annulus that is too small for any of the DMDs, as shown in the Strehl ratio map corresponding to this relay (**Figure 2d**). Use of 0.12 NA Offner relays in early semiconductor lithography (e.g. Perkin Elmer Micralign 100) was enabled by translation systems that simultaneously moved the wafer and mask across the narrow annular field[65]. This approach adds significant mechanical complexity and was estimated to be too expensive and complex for a desktop device. An Offner-like relay with aspherical mirrors was also considered, also in order to increase the diffraction-limited imaging to the full area of large DMDs **(Figure 2e)**. This approach works well according to the Strehl ratio map (**Figure 2f**), but the large aspherical mirrors were judged to be difficult to manufacture and expensive. An Offner-like catadioptric relay with two mirror surfaces and two refractive elements (**Figure 2g**) was also modeled, but results both in an excessively large system and unsatisfactory imaging along the long edges of the larger 4K DMD (**Figure 2h**). The design goals were satisfied by a 0.12 NA catadioptric relay consisting of one biconvex and one biconcave spherical lens and a concave spherical mirror (**Figure 2i**). This relay design results in diffraction-limited imaging over a field larger than any existing DMD (**Figure 2i**) and is comparable to the aspherical Offner relay, but with much smaller optics (16 cm diameters) and simpler surfaces. An important disadvantage of catadioptric optics vs an Offner relay is their use of refraction, which potentially result in chromatic aberrations with the use of polychromatic illumination. However, this relay design is well corrected for first order axial as well as lateral chromatic aberration. Thus, the relay can be used with LED sources at least between 365 and 405 nm (the most relevant range for photodeprotection of modern photolabile groups[66-67]). However, shifting between 365 and 405 nm LEDs requires a minor refocus.

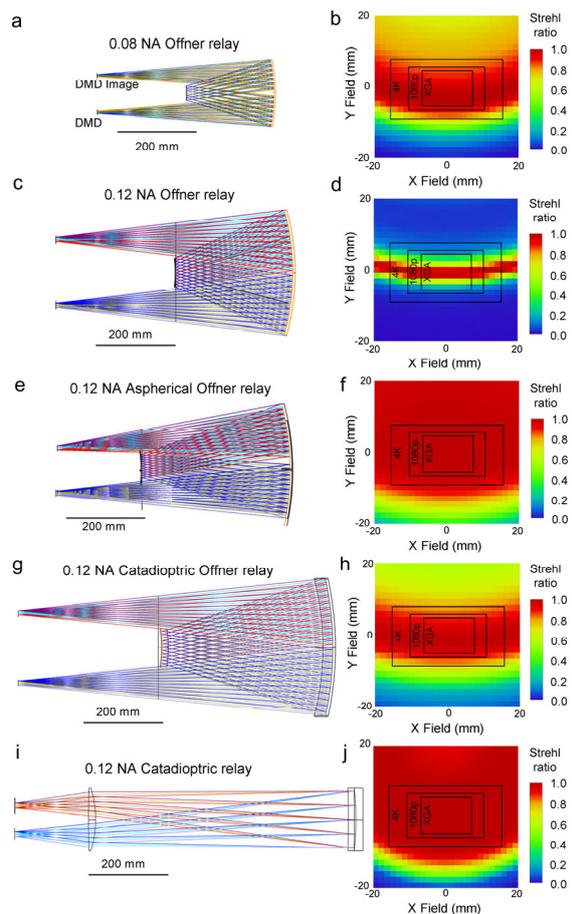

**Figure 2**. **a**. Reference 0.08 NA Offner relay of the original MAS. **b**. Strehl ratio map of the image formed by the 0.08 NA Offner relay in a 40 × 40 mm field with rectangles representing the sizes of XGA, 1080p and 4K DMDs. **c**. An Offner relay with a NA of 0.12 requires primary and secondary mirrors with about double the diameter while providing only a narrow annulus of diffraction limited imaging (Strehl ratio above 0.8) as shown in **d**. An alternative relay design exploration examined: **e**. a 0.12 NA Offner-like relay with aspherical mirrors which provides a greatly improved imaging field **f**; **g**. a 0.12 NA Offner-like catadioptric relay with four spherical surfaces yielding an improved but insufficient imaging field **h.** The optimal solution was found to be a 0.12 NA catadioptric relay **i** with relatively small diameter components yielding a Strehl ratio field **j** able to accommodate all DMD sizes

### Catadioptric relay

Having identified the catadioptric relay as the optimum choice for the next generation of desktop maskless array synthesizers, the design was further optimized, primarily to minimize the size of the relay optics while adding a fold to allow sufficient physical separation between the DMD and the DMD image. This separation is needed to accommodate the associated electrical, mechanical and fluidics systems associated with both the DMD and the photochemical reaction chamber into which the DMD image is projected. The catadioptric relay was optimized combining the single spherical mirror with a fused silica dialyte in a double pass configuration (**Figure 2i**), similar to a Dyson system[68], but with a longer working distance and an additional lens adjacent to the mirror, which is a first surface mirror and not a Mangin mirror[69]. The design results in interesting first order aberration properties. The dialyte—combined front and back lens and the large air gap between them—forms a Schupmann lens[70] that can be corrected for axial color, and the double-pass 1:1 imaging configuration remains color corrected. Additionally, due to the symmetry about the stop at the mirror, the lateral color, coma and distortion are canceled; and for sufficiently large element separation, all seven third-order aberrations are correctable using the curvature of the mirror to counteract field curvature, and the power and bending of the lenses to correct spherical aberration and astigmatism.

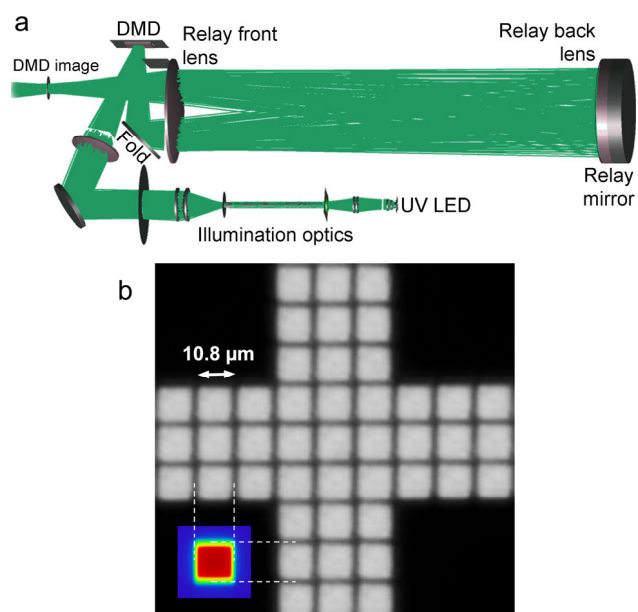

The following parameters were assessed in the optimization of the relay: maintenance of Strehl ratio performance over the spectral width of the 365 nm LED source (Nichia NWSU333B)[71], and allowing for minor refocusing, the circa 50% larger spectral width of the 405 nm variant of the same Nichia LED, depth of focus as quantified by changes to the Strehl ratio and point spread function (PSF), mechanical mounting tolerances to minimize astigmatism, scatter and stray light as detected at the DMD image, and temperature stability.

The optimized relay along with the DMD illumination optics is shown in **Figure 3a**. A fold was introduced in the relay in order to facilitate DMD illumination and to provide sufficient space for the DMD, DMD mechanical support and associated electronics, as well as to provide sufficient space for the photochemical reaction chamber into which the DMD image is projected. The image quality is shown in Figure 3b, an image of a cross of 1080p DMD mirrors captured by a camera. The shapes of the mirror images, as well as the light instantaneity of the imaged mirror gaps are consistent with diffraction-limited imaging at 365 nm with a numerical aperture of 0.12.

**Figure 3**. **a**. Full optical system of the MAS 2.0 showing the fold in the relay and the illumination system. **b**. Photo of a pattern of individual mirrors in the image plane of the catadioptric relay with an insert showing a simulated aerial image insert of an isolated single mirror in a diffraction limited NA 0.12 system.

The most effective photolabile protecting groups are *o*-nitrobenzyl derivatives[72]. These are used to protect a wide variety of functional groups[73], and primarily absorb in the near ultraviolet. 365 nm is usually chosen for the photodeprotection due to the strength of the mercury I line, and more recently, the introduction of gallium nitride-based 365 nm LEDs with indium gallium nitride ($In_xGa_{1-x}$)N quantum wells[74-75]. Compact light sources further into the UV are generally weaker and more expensive, but more importantly, the light is more strongly absorbed by the biomolecules, potentially resulting in photochemical damage that effectively reduces the synthetic yield[76]. A shift to the red edge of the UV spectrum, or to blue edge of visible light, is potentially advantageous due to reduced potential for photochemical side reactions during synthesis. The newest *o*-nitrobenzyl derivative, thiophenyl-2-(2-nitrophenyl)propoxycarbonyl (SPh-NPPOC)[66] absorbs sufficient near UV light to be useful for photolithographic DNA synthesis and it was therefore a priority to develop a relay compatible with high power LED sources ranging from 365 to 405 nm.

The catadioptric relay (**Figure 2i**) was found to satisfy this degree of achromatism as illustrated in **Figure 4**. In particular, **Figure 4a** shows the wavelength-dependent Strehl ratio over the range of the typical emission spectrum of a 365 nm Nichia NWSU333B UV LED. The Strehl ratio is near one for representative points on the image of the 1080p DMD. In the case of the emission of the 405 nm LED, the Strehl ratio is drops below 0.8 at the top center of the image of the DMD image (field 2), only for the very red edge of the spectrum, which accounts for a negligible proportion of the total irradiance. The slight favorable shift in the wavelength-dependent Strehl ratios between **Figure 4a** and **4b** result from a small adjustment of the image focus.

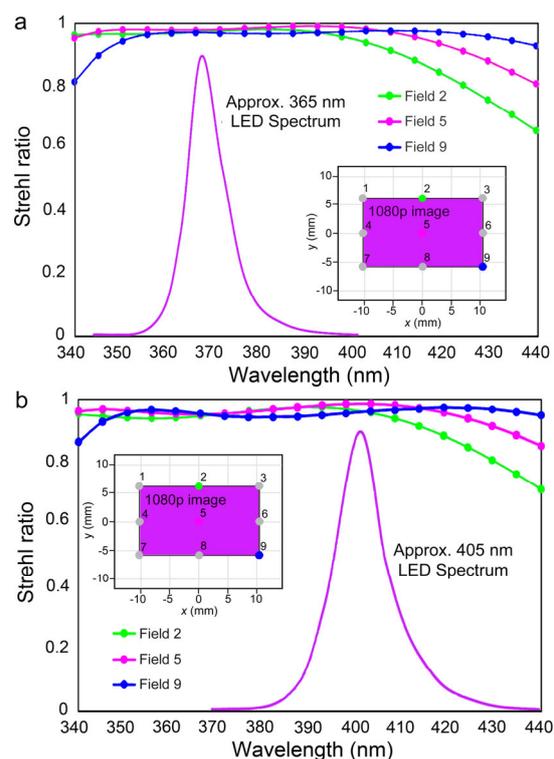

**Figure 4**. Achromaticity of the catadioptric relay in the near ultraviolet from 340 to 440 nm. **a**. Wavelength dependent Strehl ratio at three representative spots on the image of a 1080p DMD showing diffraction limited performance over the full range of a representative emission spectrum of a 365 nm high power UV LED. **b**. Wavelength dependent Strehl ratio at the same three spots. The Strehl ratio is slightly below 0.8 for field 2 for the edge of the red tail of the emission spectrum of a 405 nm LED.

**Illumination system and numerical aperture**

The illumination system is designed to efficiently collect and spatially homogenize light from either a Nichia NWSU333B 365 or 405 nm LED for imaging onto either a 0.7" XGA or 0.95" 1080p DMD with a numerical aperture up to 0.12 (**Figure 5**). A lower NA combined with the XGA DMD can be useful in microarray synthesis because the larger depth of focus (circa 50 μm vs 23 μm for 0.12 NA) allows parallel synthesis on a second substrate. For this purpose, a flow cell was designed with two synthesis substrates, separated by 50 μm, as both entrance and exit windows on the photochemical reaction chamber; after synthesis, the two substrates are both equally functional as microarrays[77].

A key consideration in photolithographic synthesis is the chemical kinetics of photodeprotection. Photodeprotection follows 1st order kinetics, meaning that the rate of removal of the protecting groups is proportional to the concentration of the protecting groups: rate = -d[$A$]/d$t$ = k[$A$], where [$A$] is the concentration of the protecting groups and k is a constant that is determined by the quantum yield for photodeprotection at a given wavelength, as well as the radiant intensity of the light used for photodeprotection. This differential equation has the solution: $[A] = [A]_0 e^{-kt}$, where $[A]_0$ is the concentration at $t = 0$. A consequence of 1st order chemical kinetics in photolithographic synthesis is that optical contrast has a strong effect on the error rate. This is because unintended exposure, due to both diffraction and scattered light, always reaches areas of the synthesis surface with the maximum concentration of protecting groups (**see Figure 1**), whereas intentional exposure reaches areas of the surface where the protecting group concentration is in the process of being depleted and is therefore always lower[57, 78]. Thus, on average, each photon of stray light removes more protecting groups than a photon reaching its intended target. Therefore, strategies to minimize stray light are necessary to maximize the yield of correct sequences.

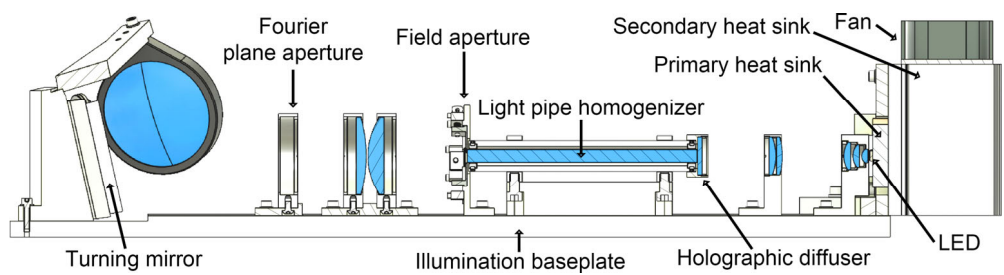

**Figure 5.** MAS illumination system. All optical components are off-the-shelf. A holographic diffuser can be placed in upstream of the light pipe homogenizer to optimize the light distribution at the Fourier plane aperture.

It is convenient to use the illumination system to increase contrast by manipulating the angular distribution of light within the numerical aperture. Similar strategies have been used in astronomy for corona imaging, as well as in the search for extrasolar planets, both of which requires high contrast to detect many orders of magnitude weaker light relative to an immediately adjacent star[79-80]. Light scattered within the imaging system, as well as the diffraction pattern, limit contrast and hence detectability of the corona or of faint solar companions. Contrast can be effectively enhanced by using apodization to create a Gaussian aperture pupil mask to replace a more typical circular aperture. The circular aperture creates an Airy diffraction pattern, the Fourier transform of the aperture. Since the

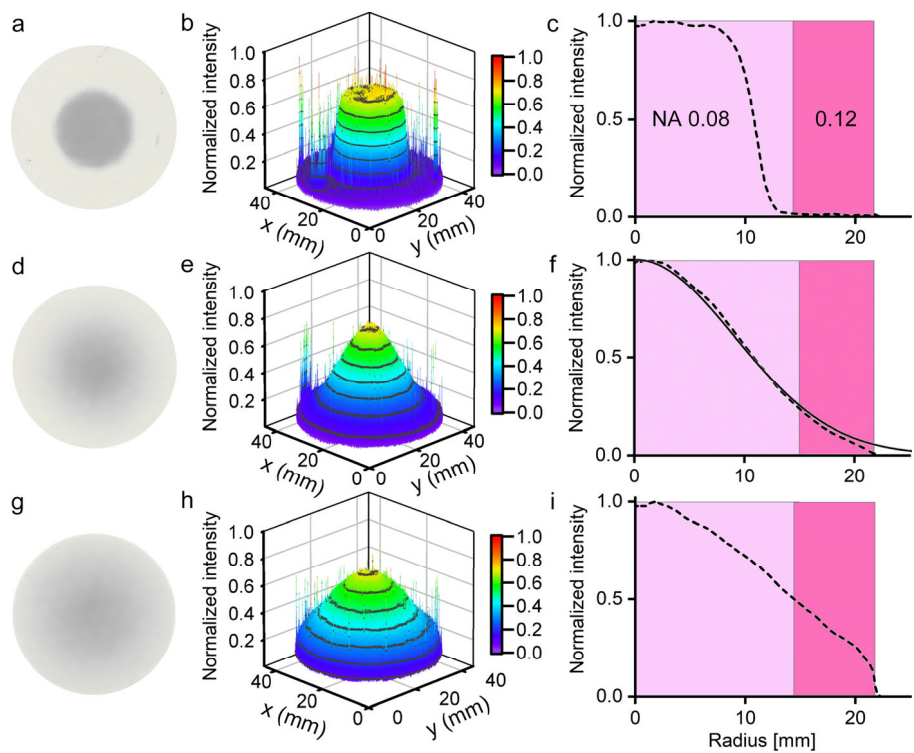

**Figure 6**. Distribution of light at the Fourier plane aperture of the illumination system. UV-sensitive radiochromic film was used the map the distribution of light within a physical aperture with a maximum NA of 0.12. **a**. Exposed radiochromic film with a diameter corresponding to a numerical aperture of 0.12 exposed without any diffusing elements. **b**. 3D representation of the film exposure (spikes are due to dust and imperfections in the film). **c**. Graph of radial intensity average over the circle with colored areas indicating the distribution needed for numerical apertures of 0.08 and 0.12. **d**. Exposed film exposed with a 10° holographic diffusor immediately before the light pipe homogenizer. **e**. 3D representation of the film exposure with 10° holographic diffusor. **f**. Graph of radial intensity average with the 10° holographic diffusor. The solid line is a Gaussian fit. **g**. Exposed film exposed with a 20° holographic diffusor immediately before the light pipe homogenizer. **h**. 3D representation of the film exposure with 20° holographic diffusor. **i**. Graph of radial intensity average with the 20° holographic diffusor.

Fourier transform of a Gaussian is also a Gaussian, a Gaussian aperture pupil results in a higher contrast image without an Airy pattern but with a lower resolution due to the loss of the high frequency components. We hypothesized that by using a holographic diffuser in the illumination system, the angular distribution of light could be made to approximate a Gaussian without the use of a complex apodization pupil.

**Figure 6** shows the distribution of light in the Fourier plane of the illumination system without a diffusing element as well as with 10° and 20° holographic diffusers to distribute low NA light to higher angles to approximate a Gaussian distribution. The following diffusers were used: 10° (Edmund Optics fused silica holographic diffuser, #48-506) and 20° (Edmund Optics fused silica holographic diffuser, #48-508). The measurements were performed by exposing radiochromic film filling the full diameter of the Fourier plane aperture. Without a diffusing filter, the NA is below 0.08 but with a distribution with a partial Gaussian character (**Figure 6a, b, c**). With a 10° holographic diffuser, there is a significant shift of intensity resulting in a distribution very close to a Gaussian fitting well into the maximum available numerical aperture of 0.12 as shown in **Figure 6d, e, f**. The dashed line in **Figure 6f** is the measured light radial distribution and the solid line is a Gaussian fit ($A = e^{-(r/w)^2}$) where $r$ is the aperture radius and the Gaussian width $w$, $1/e$, is approximately 12 mm. A 20° holographic diffuser (**Figure 6g, h, i**) further shifts light to higher angles at the illumination pupil, resulting in a higher NA, but some of the light is lost to angles corresponding to an NA greater than 0.12 and the distribution is significantly distinct from a Gaussian. The use of diffusers results in a loss of light, but the holographic diffusers are relatively efficient. Relative to the 365 nm intensity at the DMD image without a diffuser, the intensity with the 10% holographic diffuser is reduced to 86% and, with the 20% holographic diffuser is reduced to 68%, as measured with a calibrated UV intensity meter.

The result of Gaussian apodization (**Figure 6d, e, f**)—in comparison with the full illumination of the 0.12 NA aperture—can be seen in the simulated aerial images of 1080p mirror patterns corresponding to these two illumination conditions (**Figure 7**). The Gaussian apodization increases chemical-synthesis-relevant optical contrast, decreasing light intrusion into adjacent synthesis pixels. Loss of some high frequency components due to the apodization slightly increases the unintended illumination of the interstitial regions corresponding to the gaps between mirrors, but synthesis errors in these areas of the surface are generally more easily tolerated because the error rate is so high that the resulting oligonucleotides either do not participate in the experimental process, or they can be more easily separated from lower error oligonucleotides, e.g. in the sequencing data. Alternatively, the synthetic surface can be passivated on a grid pattern corresponding to interstitial regions to prevent any synthesis. The benefits of Gaussian apodization are likely subtle or undetectable in hybridization based experiments, but are likely both detectable and beneficial, for example, in data storage applications, where the higher optical contrast should be detectable in the sequencing data.

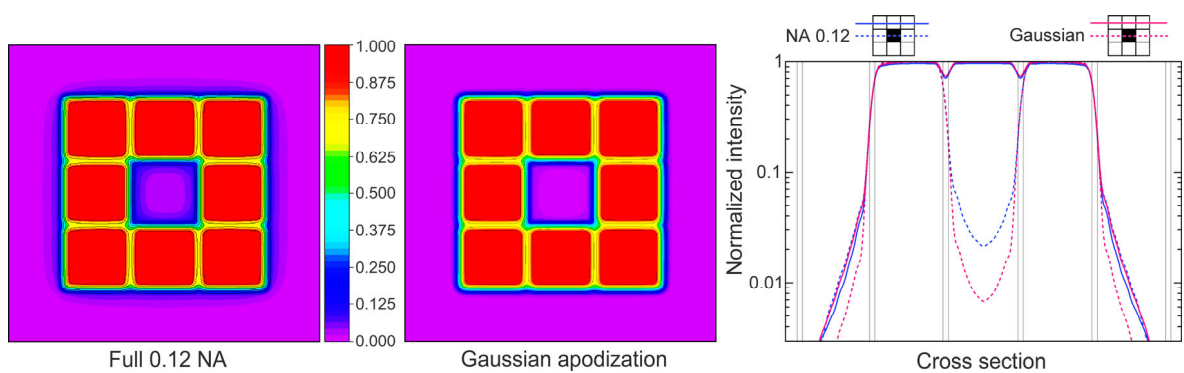

**Figure 7.** Calculated aerial images and intensity cross sections of a pattern of eight 1080p DMD mirrors imaged using a numerical aperture of 0.12 and with the Gaussian apodization described above. The mirrors are modeled as 10 × 10 µm squares on a grid with pitch 10.8 µm.

**Light scattering and its suppression**

DMD imaging maximizes contrast with a 24° separation between off-state and on-state reflected light, but this separation is reduced with increasing numerical aperture. In addition, the DMD itself scatters light, from the mirror edges, from the via connecting the mirror to the substructure, from the substructure below the mirror gaps[81]. Imperfections and dust within the relay additionally contribute to scattered light reaching the synthesis area. To maximize contrast and minimize the undesired removal photolabile groups in areas of the surface not being deliberately exposed, the scattering of light should be suppressed to maximize the yield of correct molecular sequences.

As mentioned above in the context of diffraction, photodeprotection follows first order reaction kinetics, therefore low levels of stray light reaching the DMD image in the photochemical reaction chamber will have a disproportionately large impact on the insertion error rate (unintended photodeprotection followed by addition in the next chemical coupling cycle; see **Figure 1**). Several measures were taken in the system design to suppress stray light. Relative to the original Offner relay system, the increase in numerical aperture from 0.08 to 0.12 could result in a circa 2-fold increase in stray light due to the increase in solid angle from which scattered light is directed to the image. Reflective versus refractive optics contribute differently to loss of contrast, with rough estimates suggesting that reflective surfaces contribute four to ten times more than refractive optics, with potentially large contributions originating from defects in aluminum reflective coatings[82]. Too a first approximation, and with a shared value of numerical aperture, the catadioptric relay, with light passing twice through each of two lenses, plus a single reflection, should have a similar scattering contribution to synthesis errors as an Offner relay with two reflections on the primary mirror and one on the secondary mirror. In order to mitigate an increase in scattered light due to the increase in numerical aperture, the fused silica substrates for the lenses of the catadioptric relay were specified to have few internal defects that can scatter light (Corning 7980 A1; ISO 10110 2/4;4). Also, the finish of the surfaces was held to five maximum surface imperfections of less than 0.16 mm, three long scratches shorter than 0.005 mm (ISO 10110-5 5/5×0.16; L3×0.005). In a laboratory environment, dust can accumulate on optical surfaces and increase scattering, and in this respect the catadioptric relay is likely to perform better over time than an equivalent Offner relay because the relay components can be protected from dust by using the front lens acting as window into this isolated compartment. In the case of the Offner relay, the solvents used in nucleic acid synthesis can easily reach both mirror surfaces, where they can photopolymerize on the aluminum surfaces. Thus, the Offner relay requires more frequent cleaning, and this cleaning is more likely to damage the fragile aluminum reflective coatings on the two mirrors, resulting in scratches that, along with dust, increase scattering. In order to quantify differences in scattering levels between maskless array synthesis with Offner vs catadioptric relays, we placed a UV-sensitive photodiode in the center of the image in both systems. The Offner relay-based MAS system (**Figure 2a**) has been extensively described previously[54, 78, 83]. Essentially, it consists of a concave primary mirror (diameter 185.96 mm; R450.00 +0.00/-0.30 mm) and a secondary convex mirror with half the radius of curvature (diameter 36.58 mm; R225.29±0.10 mm), with each spherical surface sharing a common origin at R=0, and provides a numerical aperture of 0.08.

To make a valid comparison, we reduced the numerical aperture of the catadioptric relay to 0.08 by using apertures in the Fourier planes of both the illumination system and the relay itself (directly in front of the back lens). Stray light was quantified by placing a UV-sensitive photodiode in the center of the DMD images of the two systems and projecting a series of DMD virtual masks patterns as shown in **Figure 8**. A circular center of OFF mirrors greater than the diameter of the photodiode (represented as a yellow circle in the figure) ensured that only scattered light is detected. In **Figure 8a**, the fraction of ON mirrors was increased in two steps from 0 to 100% and the corresponding photodiode signal was recorded. As expected, the detected light is linear with the number of ON mirrors, but in all cases, the scattered light is about 1/3$^{rd}$ higher for the Offner relay. This experiment was repeated for three radii values of the central circle of OFF mirrors as shown in **Figure 8a**. In **Figure 8b**, the number of ON mirror in a ring centered on the photodiode remains constant, but the diameter of the ring is increased stepwise from 400 mirrors to 700, measured in multiples of the 13.8 µm pitch mirrors. For both the catadioptric and Offner relays, the scattered light signal falls exponentially with distance. Fitted with the equation $I(x)=I_0+Ae^{-x/d}$, where $I$ is the scattered light intensity, $I_0$ is the intensity offset, $A$ is the amplitude, and $d$ is the distance parameter, the catadioptric relay data is well-fit (adjusted $R^2$ =0.999) with values of $I_0$ = 2.56, $A$ = 1253187, and $d$ = 47.7. For the Offner relay data, fitted to the same single exponential decline function the fit values are: $I_0$ = 3.15, $A$ = 2304424, and $d$ = 43.3 (adjusted $R^2$ =0.996). The vertical offset likely mostly originates from scattering from micromirror edges of OFF mirrors that are directly imaged onto the photodiode as well as photodiode dark current[57]. The first set of scattering experiments served to confirm the linearity of scattering, as well as the relative scattering performance of the two relay types. The second set of experiments allows a quantification of the distance dependence of scattered light. Together, these scattering results can be used to estimate the global scattering contribution to synthesis error (insertion rate) in DNA (or similar) synthesis. Local scattering, primarily scattering from the micromirrors themselves, particularly the edges and central divot in each micromirror above its support column, are imaged onto the synthesis surface and therefore contribute to errors in the direct vicinity of synthesis pixels. Synthesis errors due to local scattering, which is unlikely to be dependent on the type of optical relay, along with errors due to diffraction can be elucidated via ongoing DNA sequencing experiments. In order to test the contribution of the larger solid angle of light collection in the 0.12 catadioptric relay, the aperture corresponding to a NA of 0.08 was removed from the Fourier plane of the relay, and equivalent measurements to those shown in **Figure 8** were taken. The linearity experiments (**Figure 8a**) resulted in an increase in detected scattered light of 8.2–10.3% depending on the particular mask set. The distance dependence (**Figure 8a**) was unaffected by the change in numerical aperture. The naïve hypothesis suggests that the scattered light should be approximately proportional to the solid angles corresponding to each numerical aperture, 0.051 and 0.114 sr, for NA 0.08 and 0.12, respectively. That the factor of two in solid angle is not predictive of the increase in measured scattered light is likely partially due to the Gaussian-like apodization of the illuminations system (**Figure 6f**), which reduces light incident on areas of the second lens and mirror of the catadioptric relay corresponding to the higher NA. The tail of the Gaussian-like apodization includes about 6-7% of the total light, which accounts for most of the increase in detected scattered light.

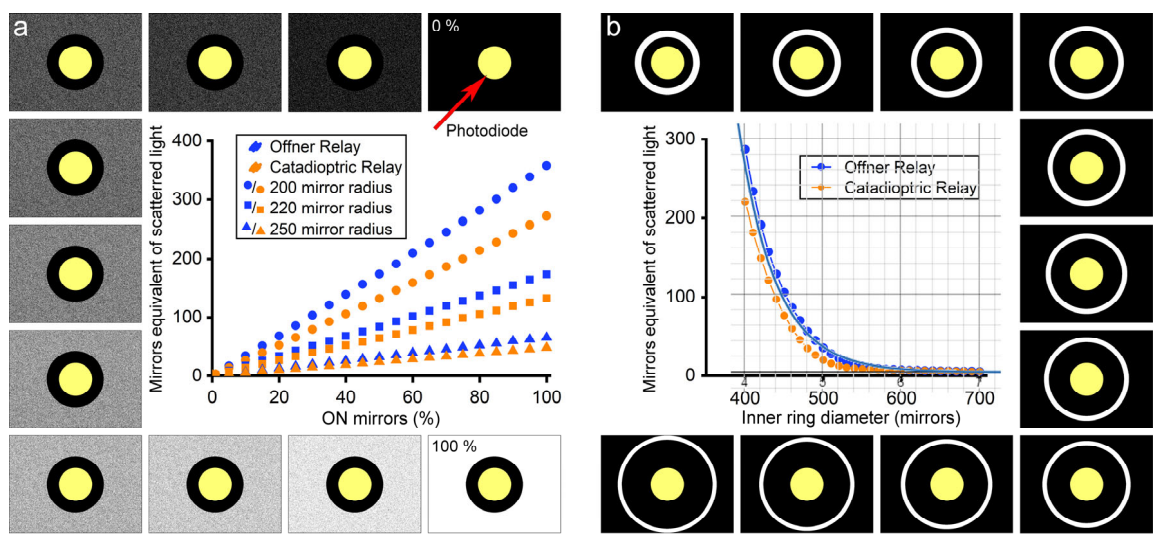

**Figure 8**. Evaluation of scattered light reaching the image in both the catadioptric and Offner relay maskless array synthesizers at a numerical aperture of 0.08 and equipped with 0.7" XGA DMDs. Scattered light was measured using a UV-sensitive photodiode placed at the center of the image plane (yellow circle). Multiple patterns of ON (white) and OFF mirrors (black) were used. **a**. comparison of the catadioptric and Offner relays using virtual masks with increasing numbers of micromirrors directing light to the image. **b**. Distance (in units of mirror pitch – 13.8 µm) dependence of scattered light in the catadioptric and Offner relays. For each ring of ON mirrors, the number of mirrors remains constant but the inner ring diameter increases. Scattered light intensity is given as units of mirror equivalents, the amount of light a single mirror would contribute to the measured signal if directly illuminating the photodiode.

**Conclusion**

A catadioptric relay system consisting with just two lenses of the same material and a single mirror has been shown to be optimal for maskless photolithographic synthesis of DNA, other nucleic acids, and more generally, any sequence-defined chemical synthesis of oligomer arrays. The catadioptric relay provides sufficient achromatism to allow the use of both 356 and 405 nm high power UV LEDs, and greatly increases the throughput of maskless array synthesis by enabling the use of modern digital micromirror devices. The design, along with detailed protocols for nucleic acid synthesis, and extensive ancillary information is available open source[83].

**Methods**

Experiments were performed on two generations of maskless array synthesizers. The new catadioptric-relay-based synthesizer (MAS 2.0, **Figure 9a**) was built as recently described[83]. The previous generation of Offner-relay based maskless array synthesizers (**Figure 9b**) has already been extensively described in the literature[54, 57, 78, 84-85].

Measurements of the distribution of light at the Fourier plane aperture of the illumination system were performed with the use of radiochromic film (Far West Technology FWT-60-20F). With 365 nm light, this radiochromic film darkens sufficiently for analysis with radiant exposures starting at about 4 J/cm$^2$. In order to quantify the light intensity distribution, we first calibrated the linearity of the film darkening. For this, a calibrated UV light intensity meter (Ushio Unimeter UIT-201) with a 365 nm probe was used to measure the intensity of UV light at the DMD image position with all micromirrors tilted to the ON position. With radiochromic film cut to shape and inserted into the flow cell, a series of square mirror patters were displayed on the DMD for increasing time intervals in order to expose the film to ten radiant exposures between 2.1 and 21 J/cm$^2$ as shown in **Figure 10**. After the exposure, the radiochromic film was scanned (**Figure 10a**) and the degree of film darkening was extracted using ImageJ (**Figure 10b**). In this range, the calibration demonstrates that the color change is linear with radiant exposure (**Figure 10c**). In subsequent exposures in the illumination system, the film was cut to a circle to fit the aperture holder and exposed within this linear range. The film was then scanned and the color change information extracted using ImageJ (**Figure 6**).

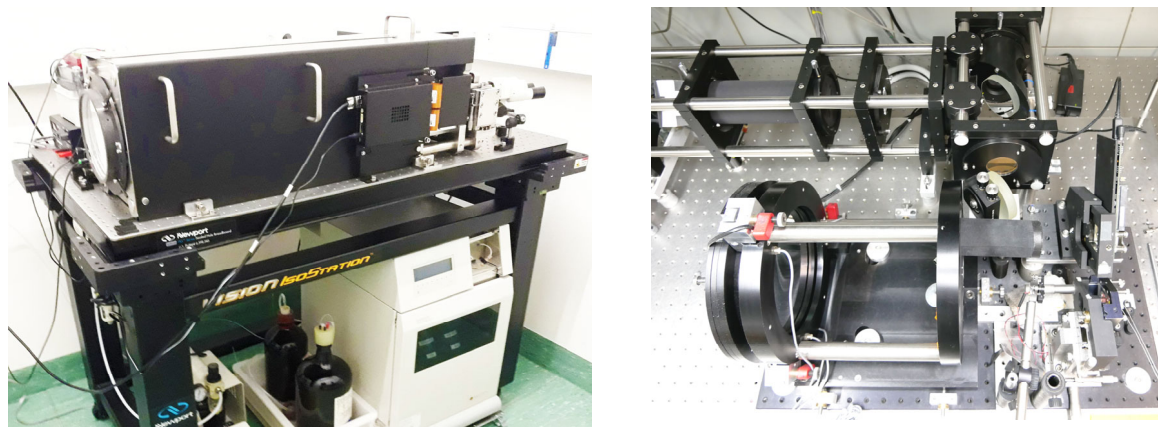

**Figure 9**. **a**. Catadioptric relay maskless array synthesizer with the chemical delivery system, an Expedite 8909 nucleic acid synthesizer, underneath. **b**. Offner relay maskless array synthesizer with dust covers removed. The primary mirror of the Offner relay is visible in the left foreground.

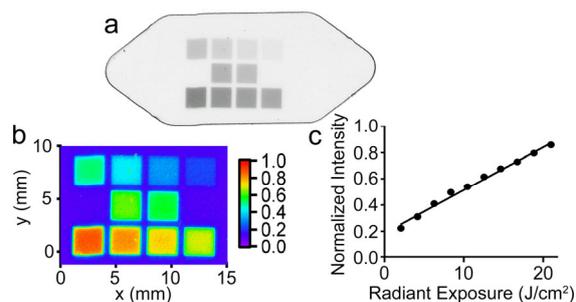

**Figure 10**. **a**. Image of radiochromic film with square features corresponding to ten different radiant exposures. The irregular hexagonal shape of the film corresponds to the dimensions of the flow cell used for DNA synthesis **b**. ImageJ representation of the color change due to exposure. **c**. Graph of radiant exposure versus intensity of the color change with a linear fit of the data.

Scattering quantification in both the catadioptric- and Offner relay-based synthesizers was accomplished by using a 3D-printed adapter to position a UV-sensitive photodiode (sglux TOCOM_A5) at the center of the DMD image. Illuminated by the 365 nm UV LED (Nichia NWSU333B) via the respective illumination systems, the DMD (Texas instruments 0.7" diagonal UV XGA DMD; 13.68 µm micromirror pitch) was used to project a series of patterns as shown in **Figure 8**. Each of the patterns—virtual masks—was a 1 bit bitmap XGA image, with the black pixels resulting in a tilt of the corresponding micromirror to the OFF position and the white pixels corresponding to micromirrors tilted to the ON position. The center of each virtual mask had a central circle of black pixels to prevent direct illumination of the photodiode. Three values of the radius of this circle were used: 200, 220, and 250 pixels. In the first series of scattering measurements (**Figure 8a**), the pixels outside the black circle were filled with a random mix of black and white pixels such that each mask in the series had an increasing proportion of white pixels, spanning the range from 0% white to 100% white in 20 steps. These virtual masks were then used to project the pattern of UV light onto the image plane and the photodiode signal recorded with a voltmeter. The scattering signal intensity was calibrated using a series of virtual mask with only a few, central white pixels resulting in the direct illumination of the photodiode. Thus, the scattering is given in units of mirror-equivalents of light intensity, a value which is independent of the LED illumination intensity. In the second series of scattering experiments, the virtual masks consisted of a ring of white pixels centered on the position of the photodiode. The inner diameter of the ring ranged from 400 pixels (mirrors) to 700 pixels in 30 steps, and the width of the ring was reduced in each successive mask such that all rings consisted of the same number of white pixels. The same intensity calibration as before was used for these experiments.

**Funding**

The authors gratefully acknowledge financial support from the German Research Foundation (DFG) under grant 460736965, from the European Union's Horizon Europe Research and Innovation Program under grant agreement 101070589, and from the European Innovation Council Pathfinder Challenge: DNA-based digital data storage under project number 101115134.